\documentclass[10pt]{wlscirep}
 \pdfoutput=1 
\usepackage[utf8]{inputenc}
\usepackage[T1]{fontenc}
\usepackage{graphicx,amsbsy,amsmath,amsfonts}
\usepackage{amssymb,amsmath,epsfig,stmaryrd}
\usepackage{longtable}
\usepackage{todonotes}

\def\ds{\displaystyle}

\newcommand{\bu}{{\bf u}}
\newcommand{\bG}{{\bf G}}

\newcommand{\mbx}{\mathbf{x}}
\newcommand{\mbl}{{\boldsymbol\ell}}
\newcommand{\bn}{\bf n}

\title{Spatio-temporal predictive modeling framework for infectious disease spread}
\author[1]{Sashikumaar Ganesan}
\author[1]{Deepak Subramani}
\affil[1]{Department of Computational and Data Sciences, IISc Bangalore \newline sashi@iisc.ac.in, deepakns@iisc.ac.in}

\begin{abstract}
A novel predictive modeling framework for the spread of infectious diseases using  high-dimensional partial differential equations is developed and implemented. A scalar function representing the infected population is defined on a high-dimensional space and its evolution overall directions is described by a population balance equation (PBE). New infections are introduced among the susceptible population from a non-quarantined infected population based on their interaction, adherence to distancing norms, hygiene levels and any other societal interventions. Moreover, recovery, death, immunity and all aforementioned parameters are modeled on the high-dimensional space. 
To epitomize the capabilities and features of the above framework, prognostic estimates of Covid-19 spread using a six-dimensional (time, 2D space, infection severity, duration of infection, and population age) PBE is presented. Further, scenario analysis for different policy interventions and population behavior is presented, throwing more insights into the spatio-temporal spread of infections across duration of disease, infection severity and age of the population. These insights could be used for science-informed policy planning. 
\end{abstract}
\begin{document}

\flushbottom
\maketitle
%
%
\thispagestyle{empty}
 
\section*{Introduction}
Epidemic modeling and forecasting has gained renewed interest since late 2019 when the world was affected by the novel coronavirus pandemic (named Covid-19). Several computational studies to predict the human-to-human spread of Covid-19 have been reported~ \cite{singh2020age,harsha2020covid,pandey2020seir,ranjan2020predictions}. Most of these efforts have been based on compartmental models and stochastic models (including agent-based models)  \cite{vynnycky2010introduction}. In compartmental models (e.g., SIR, SEIR, SEIRS, DELPHI~ \cite{vynnycky2010introduction,mitdelphi}), the population is divided into different compartments and the dynamics of the different compartments are modeled by a system of coupled ordinary differential equations (ODE). Here, the interaction among compartments is usually deterministic, whereas random processes are used to model the spread of infections in stochastic models. Agent-based models are stochastic models that undertake a bottom-up approach of modeling individual members of a population and the dynamics of their interaction in terms of probabilities of movement and contact. 

More than the total number of infections, it is essential to have more insightful predictions, e.g., infected population distribution across their age and level of infection severity for science-informed policy intervention and public health planning. The population distribution over the duration of infection is crucial for planing antiviral treatments, quarantine, ventilator support and contact tracing. This requirement necessitates a comprehensive and computationally efficient predictive modeling framework. Even though these features could be incorporated in ODE-based compartmental and stochastic agent-based models, it is very complex and computationally expensive. To overcome these challenges, we propose a novel partial differential equation-based spatio-temporal predictive modeling framework for forecasting the spread of infectious disease in heterogeneous populations in open geographies. The roots for our model lie in the population balance equations that are popular in chemical engineering and process studies  \cite{ramkrishna2002population}. 

In the proposed model, the infected population density is defined as a scalar field on a high-dimensional space. Specifically for predicting the spread of Covid-19, a six-dimensional model is presented. The first three dimensions are the space and time, and the other three are the infection severity, duration of the infection (i.e., time since infection), and age of the population. New infections, impact of quarantine, testing, contact tracing, immunity, intervention policy impact, health infrastructure, recovery, and death are all modeled on this six-dimensional space based on data-driven functions (where available), and/or simple algebraic and integral functions. 
Notably, our PDE-based model in the present paper is more compact and a versatile description of the spread of the disease compared to compartmental models, and computationally efficient compared to agent-based models. To showcase the capability of our distribution-based predictive modeling framework for infectious disease spread, we apply it to model and predict the spread of Covid-19 in India. Further, we present a scenario analysis, which could be used to draw insights for policy interventions.

\section*{Results}

\subsection*{The Population Balance Model}
Let $T_\infty$ be a given final time and $\Omega:=\Omega_x\otimes\Omega_\ell$ be the computational domain of interest. Here, $\Omega_x\subset\mathbb{R}^2$ is the spatial domain defining the geographical region of interest and $\Omega_\ell\subset\mathbb{R}^n$, where $n$ is the number of internal directions. Each of the $n$-internal directions represents the property of the population on which a distribution needs to be predicted. Suppose the properties of interest are the infection severity, duration of the infection and age of the population, then a model with three internal directions could be used as follows. Let  $\Omega_\ell:=L_v\times L_d\times L_a$ be the internal domain, where $L_v=[0, 1]$ denotes the infection severity interval, $L_d=[0,d_{\infty}]$ denotes the duration of infection, $d_{\infty}$ is the maximum duration of infection,  $L_a=[0, a_{\infty}]$ denotes the age interval and $a_{\infty}$ is the maximum age of the population. The infection index $\ell_v\in L_v$ quantifies the severity of the infection among the infected population. Specifically, the population with infection index $\ell_v=0$ is completely disease-free, with $\ell_v = 1$ has maximum severity, with $\ell_v\ge v_{\textrm {sym}}$ shows symptoms and those with $\ell_v<v_{\textrm {sym}}$ are asymptomatic. The duration of infection index $\ell_d\in L_d$ quantifies the time since a population has been exposed to and contracted the disease. Specifically, the population that just contracted the disease has $\ell_d=0$. Typically, a person is asymptomatic until they reach $\ell_v=v_{\textrm {sym}}$, and the duration elapsed $\ell_d$ is the incubation period in which the disease is sub-clinical and that population is actively spreading the disease. After recovery, a population doesn't necessarily go to $\ell_v=0$, rather they reach $\ell_v<v_{\textrm reco}$.

Let $I(t,\mbx,\mbl)$, where $t\in (0,T_\infty],~ \mbx \in \Omega_x$ and $\mbl \in \Omega_\ell$, be the infected number density function of the population.  
To describe the evolution of the active infected population size distribution, we propose the population balance equation  in the time interval $(0,T_\infty]$
\begin{equation}\label{model}
 \displaystyle\frac{\partial I}{\partial t} + \nabla\cdot(\textbf{u} I) +\nabla_\ell\cdot(\textbf{G} I) + CI = F  \quad {\textrm  {in} }  \quad \Omega_x\times\Omega_\ell\,,
\end{equation}
with initial conditions
\begin{equation}\label{model2}
 \begin{array}{rcll}
 I(t,\mbx,\mbl) &=&g_n    &{\textrm {in} }  \quad  \partial\Omega^{-}_{ x}\times \Omega_\ell\,,\\
 I(t,\mbx,(\ell_v,0,\ell_a))  & =& B_{\textrm nuc} \quad  &{\textrm {in} }  \quad  \Omega_{x}\times L_v\times L_a \,,\\
 I(t,\mbx,(0,\ell_d>0,\ell_a))  & =& 0 \quad  &{\textrm {in} }  \quad  \Omega_{x}\times L_d\times L_a \,, \\
 I(0,\mbx,\mbl) & =& I_0 &{\textrm {in} }  \quad  \Omega_{x}\times\Omega_\ell \,.
\end{array}
\end{equation}
Here, $\bu$ denotes the advection vector that quantifies the multiscale spatial movement of the population in a differential neighbourhood of $\Omega_{x}$ (e.g., migrant laborers, daily commute for work, logistics-related travel, periodic gathering for religious and social events), $\bn$  is the outward unit normal vector to $\Omega_x$, $\partial{\Omega}^{-}:= \{\mbx \in \partial{\Omega_{x}} \  |~ \bu\cdot \bn <0 \  \}$,
$g_n$ is the flux that quantifies the net addition of the infected population into $\Omega_{\textrm x}$ from outside (the spatial movement of the population across the border of the domain $\partial\Omega_{\textrm x}$), and $I_0$ is the initial distribution of the infected population.
Further,
$
\bG = (G_{\ell_v}, G_{\ell_d}, G_{\ell_a})^T
$
is the internal growth vector, where 
\begin{gather}
  G_{\ell_v} = \frac{d \ell_v}{d t} = G_{\ell_v}(\ell_a, \beta, \gamma(\ell_a), \alpha(\mbx) ),  \quad
  G_{\ell_d} = \frac{d \ell_d}{d t} = 1, \quad
  G_{\ell_a} = \frac{d \ell_a}{d t} = 1.
\end{gather}
Here, $\beta$ is the immunity of the infected population, $\gamma$ is the pre-medical history of the infected population and $\alpha$ is the effective treatment index.  Next, we define the rate term
$
 C =  C_R + C_{ID},
$
where $C_R(t,\mbx,\mbl)$ is a recovery rate function that quantifies the rate of recovery of the population from the infection, 
and $C_{ID}(t,\mbx,\mbl)$ is the infectious death rate. 
We also define a source term $F = C_T(t,\mbx,\mbl)$ that quantifies the point-to-point movement of infected population (e.g., by air, train etc) within $\Omega_x$, which are not included in $\bu$ and $g_n$.
Moreover,  $C_T$ and $\bu$ need to be defined in such a way that the net internal movement of infected population within $\Omega_{x}$ is conserved.
Moreover, $B_{\textrm nuc}$ is the nucleation function that quantifies the infection transmission from the infected to the susceptible population and it is a function of several parameters as follows
\begin{equation}
  B_{\textrm nuc} = B_{\textrm nuc}\left(t, \textbf{X}, \sigma, H, S_D, N_S, N_Q, I\right).
\end{equation}
Here, $\textbf{X}\in\Omega$, $\sigma, H$ and $S_D$ are the interactivity, hygiene and social distancing indices respectively. 
Finally, the total population $N(t)$ at a given time $t\in(0,T_\infty]$ is defined by
\begin{align*}
  N(t) &= N_S(t) + N_R(t) + N_I(t)  + N_Q(t) - N_{ID}(t) + N_B(t) - N_D(t),  \quad
  N_Q(t)  = \int_{\Omega}  \gamma_Q(t,\mbx,\mbl)I(t,\mbx,\mbl) \,dX,~\\
 N_I(t)  &= \int_{\Omega} I(t,\mbx,\mbl)\,dX, \quad
 N_{R}(t)   =  \int_{\Omega} C_RI(t,\mbx,\mbl)\,dX,\quad
 N_{ID}(t)   =  \int_{\Omega} C_{ID}I(t,\mbx,\mbl)\,dX.
\end{align*}
Here, $N_S$, $N_B$, $N_R$, $N_I$, $N_Q$ $N_{ID}$ and $N_D$ are the number of susceptible, newborn, recovered, infected (symptomatic/asymptomatic), quarantined, infectious death and natural death populations, respectively. The given initial and boundary conditions and the above defined parameters close the population balance system.

\subsection*{Modeling of Parameters}
The proposed population balance model~\eqref{model} is comprehensive and built on the basis of several parameters as defined above. In this section, we describe the modeling of each parameter.  

\subsubsection*{Nucleation}
The nucleation term $B_{\textrm nuc}$ quantifies the new infection number density that is added to the system at $\ell_d=0$ for all $t$, $\mbx$, $\ell_v$, and $\ell_a$. 
Depending on how the susceptible population interacts with the infected population, new infections are added to the system. We call this addition as \textit{nucleation} (borrowing the terminology from process engineering), which is modelled as
\begin{align}
	B_{\textrm nuc} &=  R\int_{\Omega_\ell}[1-\gamma_Q]I(t,\mbx,\mbl)\,d\ell, \label{eq:Bnuc_eq2}\\
	R &= R_0  f_1(t,\sigma)f_2(t,H)f_3(t,S_D)f_4(t,\mbx,\ell_a)f_5(\ell_v)\,,\label{eq:Bnuc_eq3}\\
	\gamma_Q &= \frac{1}{1+\exp\left(-(\ell_v-v_{\textrm {sym}})/b_v\right)} 
	  \frac{1}{1+\exp\left(-(\ell_d-d_{\textrm {sym}})/b_d\right)} 
	  \frac{1}{1+\exp\left(-(\ell_a-a_{\textrm risk})/b_a\right)}. \label{eq:Bnuc_eq4}	
\end{align}
Here, $\gamma_Q\in[0,1]$ in~\eqref{eq:Bnuc_eq2} determines the fraction of the infected population in quarantine and it can be modeled as in equation~\eqref{eq:Bnuc_eq4}. Further,  the factor $\gamma_Q$ is dependent on the level of screening including testing, strictness of enforcing isolation and compliance of susceptible general public. Suppose $\gamma_Q = 1$, i.e., if the entire infected population is kept under strict isolation, newly infected population will be zero and eventually there will be no spread of disease. However, due to economic, social and democratic reasons, implementing such a strategy is nearly impossible and there is bound to be spread, i.e., $\gamma_Q<1$. 
Moreover, the integral on the right-hand side of equation~\eqref{eq:Bnuc_eq2} is the total non-quarantined number density of the infected population at $(t,\mbx)$, and 
$R$ is the rate at which the non-quarantined population infects the susceptible population.
The factor $R$ is modelled as in equation~\eqref{eq:Bnuc_eq3}, where $R_0$ is the basic reproduction rate,
\begin{align*}
    f_1(t,\sigma) &= \left[\frac{1}{1+\exp\left(-(\sigma(t)-\sigma_c)/b_\sigma\right)}\right], \quad
    f_2(t,H)  =  \left[1-\frac{1}{1+\exp\left(-(H(t)-H_c)/b_H\right)}\right], \\
    f_3(t,S_D) &=  \left[1-\frac{1}{1+\exp\left(-(S_D(t)-S_{D_c})/b_{S_D}\right)}\right],\quad
    f_4(t,\mbx,\ell_a)  =  a_4\exp\left(-\frac{(\ell_a-b_4)^2}{c_4^2}\right), \\
    f_5(\ell_v) &=  \left\{ \begin{array}{cc}
       3\sqrt{\frac{2}{\pi}}\exp\left({\frac{-(\ell_v-v_{\textrm {sym}})^2}{2(v_{\textrm {sym}}/3)^2}}\right) & 0\leq v<v_{\textrm {sym}} \,\\
       3\sqrt{\frac{2}{\pi}}\exp\left({\frac{-(\ell_v-v_{\textrm {sym}})^2}{2((1-v_{\textrm {sym}})/3)^2}}\right)  & v_{\textrm {sym}}\leq v \leq 1.
    \end{array}   \right.
\end{align*}

\noindent Here, the interactivity index $\sigma\in[0,1]$, hygiene index $H\in[0,1]$, and social distancing index $S_D\in[0,1]$. Suppose $\sigma=0$ then everything is under perfect lockdown and $R \rightarrow 0$. In case $S_D=1$, everyone is following perfect social distancing and $R \rightarrow 0$. Moreover, the newly infected population has to be added at different age ($\ell_a$) and  infection $(\ell_v)$ levels for which the factors $f_4$ and $f_5$ are introduced. 
We propose to use logistic functions fitted to data from literature for $f_1,~f_2,~f_3$,; the normalized demography at $(t,\mbx)$ for $f_4(t,\mbx,\ell_a)$, and a Gaussian mixture with two components so that maxima is at $v_{\textrm {sym}}$ and tails are proportional to the interval length over $[0,v_{\textrm {sym}}]$ and $[v_{\textrm {sym}},1]$ for $f_5(\ell_v)$. 
In addition, the constant in $f_5$ is chosen such that the integral of $f_5$ over its support is one. This condition is imposed to ensure that $R_0$ can be interpreted as the basic reproduction rate used in standard epidemiological models  \cite{grassly2008mathematical}. 
The parameters $v_{\textrm {sym}}$, $d_{\textrm {sym}}$, $a_{\textrm risk}$, $b_v$, $b_d$, $b_a$, $b_{\sigma}$, $b_H$, $b_{S_D}$, $\sigma_c$, $H_c$, $S_{D_c}$ can be estimated from experimental and clinical evidence. Furthermore, in light of new evidence, the functional forms of $f_1$ to $f_5$ can easily be modified. Finally, $\sigma(t)$, $S_D(t)$ and $H(t)$ change over time due to increased awareness, government measures and compliance by people. 
\subsubsection*{Growth Factor}
The growth factor $G_{\ell_v}$ quantifies how the infected number density is advected along the direction of $l_v$, that is, how the infection becomes  mild to severe/critical and vice-versa in the infected population. We can model it as a function of the medical history, immunity of the population, which in turn are functions of the age $l_a$, treatment and socio-economic status. Nevertheless, as a simple first order model, we propose a nonlinear function of the age, 
\begin{align}
G_{\ell_v}(\ell_a) = K_g(\ell_a-a_{\textrm risk})^p\,,
\end{align}
where $K_g$ is a non-dimensionalization factor, $p$ is a power of nonlinearity and $a_{\textrm risk}$ is the age offset.

\subsubsection*{Recovery Rate and Infectious Death Rate}
In general, the recovery rate $C_R$ and infectious death rate $C_{ID}$ depend on $\ell_v$, and in turn are functions of hospital facilities, age, and health state of the population. These rates can be modeled directly from clinical data for all ordinates $\ell_v,~\ell_d,~\ell_a$. For the exact functional forms refer to the Supplementary Information Appendix.



\subsubsection*{Initial Infection Number Density}
The initial number density $I_0(\mbx,\mbl)$ can be estimated directly from available official data at the day of starting the simulation. However, the data is available only in-terms of total number of tested and confirmed cases at a $\mbx$-location and the dependence on $\mbl$ needs to be estimated via appropriate data-driven and analytical functions. As such, first we utilize data from a period of 14 days, along with the log-normal distribution of incubation period  \cite{lauer2020incubation} to calculate the initial number density $N_D(\mathbf{x})$ at all the spatial points $\mathbf{x}$, but integrated over the three internal ordinates ($\ell_v,\ell_d,\ell_a$), i.e.,
 \begin{equation}
     N_D(\mathbf{x}) = \sum_{i=1}^{i=14} N_i \frac{1}{i a_2\sqrt{2\pi}}\exp\left(-\frac{(\log i-b_2)^2}{2a_2^2}\right)\,, \label{eq:NDxdist}
 \end{equation}
 where $N_i(\mathbf{x})$ is the data of tested and positive. For the distribution along the internal ordinates, we propose to use the following initial infection number density distribution
\begin{align}
    I_0 &=
    N_D(\mbx)[f_5(\ell_v)]\left[a_1\exp\left(-\frac{(\ell_a-b_1)^2}{c_1^2}\right)\right] 
     \left[\frac{1}{\ell_d a_2\sqrt{2\pi}}\exp\left(-\frac{(\log(\ell_d)-b_2)^2}{2a_2^2}\right)\right]. \label{eq:I0dist}
\end{align}
Here, the first term in the square brackets is the normalized demography function (same as $f_4$), second term is the log-normal incubation period function with fitted \cite{lauer2020incubation} $a_2=0.42$ and $b_2=1.62$, and $f_5$ is same as before.

\subsection*{Covid-19 Epidemic Spread Predictions}
To exhibit the capabilities of the proposed model, the forecast of Covid-19 spread in India is presented here. The numerical scheme and the fitted model parameters are given in the Supplementary Information Appendix. The proposed model and numerical schemes are implemented in our in-house finite element package \cite{PAR01,PAR02} and have been verified in our earlier studies with applications to process engineering \cite{GAN13,GANW10}.  

With the spread of Covid-19 in India, the federal government imposed a nation-wide lockdown from March 25, 2020. To simulate the spread of infections starting from March 23, 2020, the initial distribution of infected population is estimated using the data of active cases from March 23 to April 5 according to equations~\ref{eq:NDxdist} and \ref{eq:I0dist}. Then the infection spread forecast for one year is computed by solving the PBE system (equation~\ref{model}). Further, data until June 21, 2020 is utilized to select the parameters (e.g., $S_D$, $C_R$, $C_{ID}$, $\gamma_Q$) that best explains the actual data. Thereafter, the control parameter $S_D$ is varied to perform scenario analyses as presented next.  

\subsubsection*{Scenario Analysis}
\begin{figure}[tb!]
\begin{center}
\unitlength5mm
\begin{picture}(12,13)
\put(6.5,6.75){\makebox(0,0){\includegraphics[clip, trim=0 0 0 50, width=18.5cm]{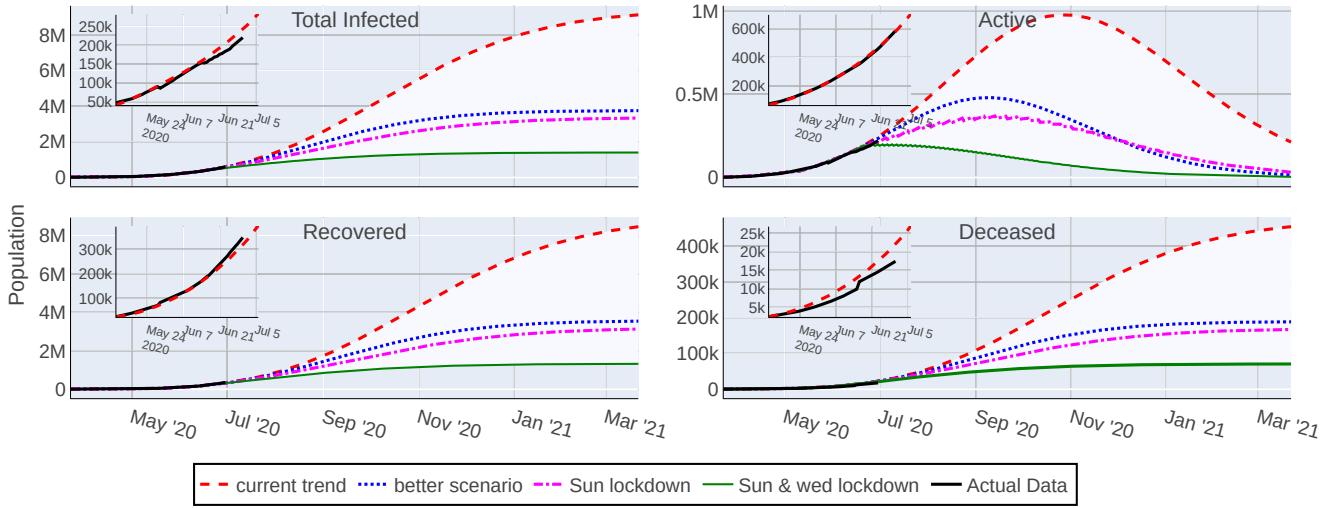}}}
\end{picture}
\end{center}
\caption{Time series forecast of active, total infections, recovered and deceased cases of Covid-19 in India from Mar 23, 2020 to Mar 22, 2021. The inset shows a zoom with comparison of the model forecast with the data until July 01, 2020 \label{fig:TimeSeriesIND}}
\end{figure}

Different future scenarios are predicted by varying $S_D(t)$ based on the anticipated individual behavior (social distancing, hygiene practice, compliance to government rules etc.) and government policies (quarantine rules, lockdown rules etc.). The first scenario, named \textit{Current Trend} follows business as usual assuming further relaxation to lockdown rules. A second variant named \textit{Better Scenario} assumes better compliance in the social distancing and other measures to control the spread of the disease. Sunday, and Sunday \& Wednesday lockdowns are imposed on the \textit{Current Trend} scenario to formulate the third and fourth scenarios respectively. These lockdown scenarios are introduced to measure the impact of periodic lockdowns on the effectiveness of these strategies to control the disease spread. 
The active ($N_I$), recovered (cumulative $N_{R}$), deaths (cumulative $N_{ID}$) and total (sum of active, recovered and deaths) predicted by the four scenarios for the duration between March 23, 2020 and March 22, 2021 are shown as time-series plots in Fig.~\ref{fig:TimeSeriesIND}. In the \textit{Current Trend}, a peak of 0.975 million ‘Active Cases’ is predicted in the last week of October 2020, and there will be around 21 million ‘Active Cases’, 450,000 deaths and 9.1 million total cases at the end of March 2021. The peak of the \textit{Better Scenario} is predicted in the second week of September 2020 with 0.478 million 'Active Cases', which is lower than the \textit{Current Trend}. Further, there will be around 14,200 ‘Active Cases’, 0.188 million deaths and 3.74 million total cases at the end of March 2021.
The weekly lockdown scenarios assume that a complete lockdown is imposed on Sunday or Sunday and Wednesday. During this lockdown, there is a complete restriction of people's movement similar to the nationwide lockdown imposed between Mar 25 and April 14 in India. With Sunday Lockdown, a peak of 0.365 million ‘Active Cases’ is sustained for about two weeks during 5-20 September 2020, and there will be around 30,200 thousand ‘Active Cases’, 0.167 million deaths and 3.32 million total cases at the end of March 2021. With Sunday and Wednesday lockdown, a peak of 0.197 million 'Active Cases' is sustained for the period 27 June to 15 July 2020, and there will around 2,800 ‘Active Cases', 70,300  deaths and 1.39 million total cases at the end of March 2021. The insets in each panel of Fig.~\ref{fig:TimeSeriesIND} show the comparison with actual data and thereby validate the model. In addition, the time series plots for other scenarios including a worse-case scenario can be found at IISc-Model website \cite{IIScModel}.

\begin{figure}[tb!]
\begin{center}
\unitlength5mm
\begin{picture}(12,8)
\put(6.5,4.25){\makebox(0,0){\includegraphics[clip, trim=0 0 0 50, width=18.5cm]{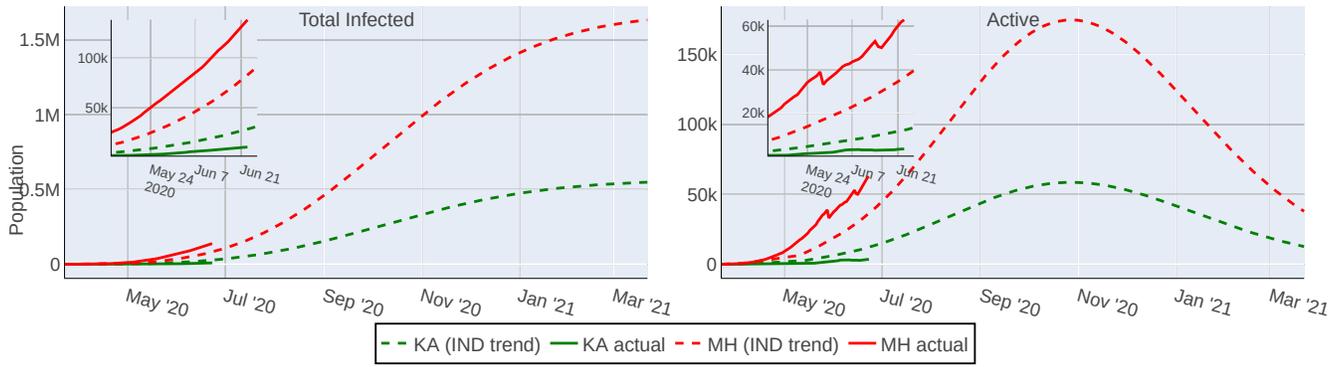}}}
\end{picture}
\end{center}
\caption{The actual data and the predictions computed with the national trend based parameters for the states of Karnataka and Maharashtra. in India. \label{fig:recovisoMH}}
\end{figure}

In order to compare the performance of all states in India with the national trend, a uniform set of parameters is used for state-wide computations. In particular, the parameters are fitted by minimizing the error between the national data and the sum  of the respective state predictions. 
Fig.~\ref{fig:recovisoMH} shows the actual data and the computed distribution using the above set of  parameters for the states of Karnataka and Maharashtra. We can see that Karnataka has done better, whereas Maharashtra has done worse compared to the national trend. These insights can be used by the authorities to introduce state-wise lockdown policies and to plan infrastructure for quarantine, treatments etc. The performance of  other states can be seen at IISc-Model website \cite{IIScModel}.

\subsubsection*{Population distribution}
\begin{figure}[t!]
\begin{center}
\unitlength5mm
\begin{picture}(12,18.5)
\put(-6,3){\makebox(0,0){\includegraphics[clip, trim=0 0 0 200,width=5.5cm]{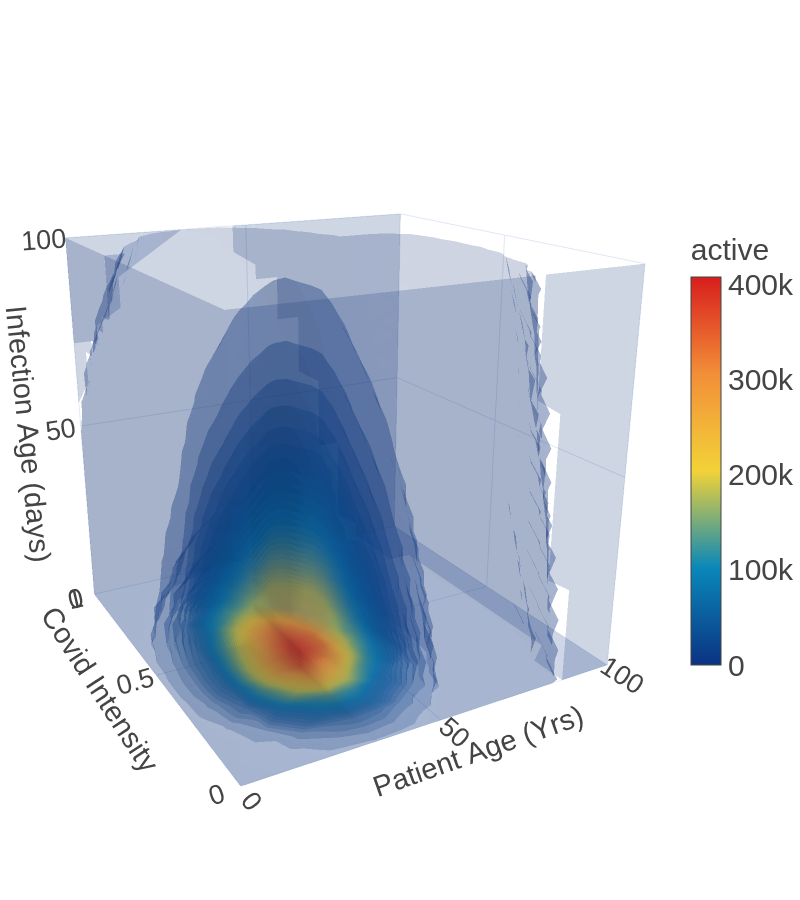}}}
\put(6,3){\makebox(0,0){\includegraphics[clip, trim=0 0 0 200,width=5.5cm]{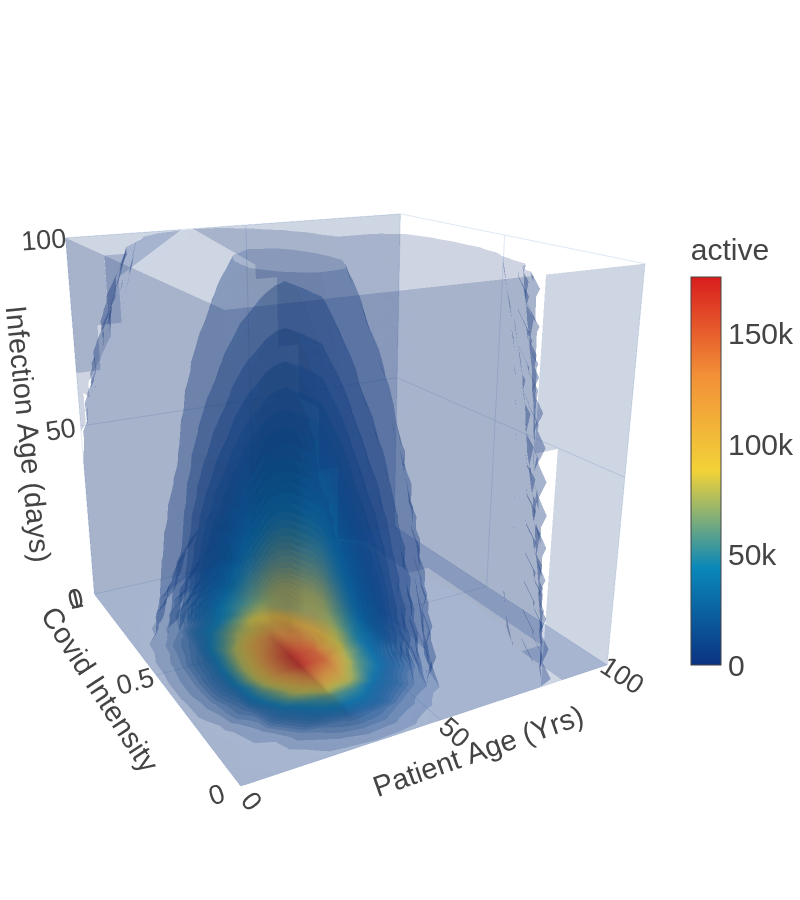}}}
\put(17.5,3){\makebox(0,0){\includegraphics[clip, trim=0 0 0 200,width=5.5cm]{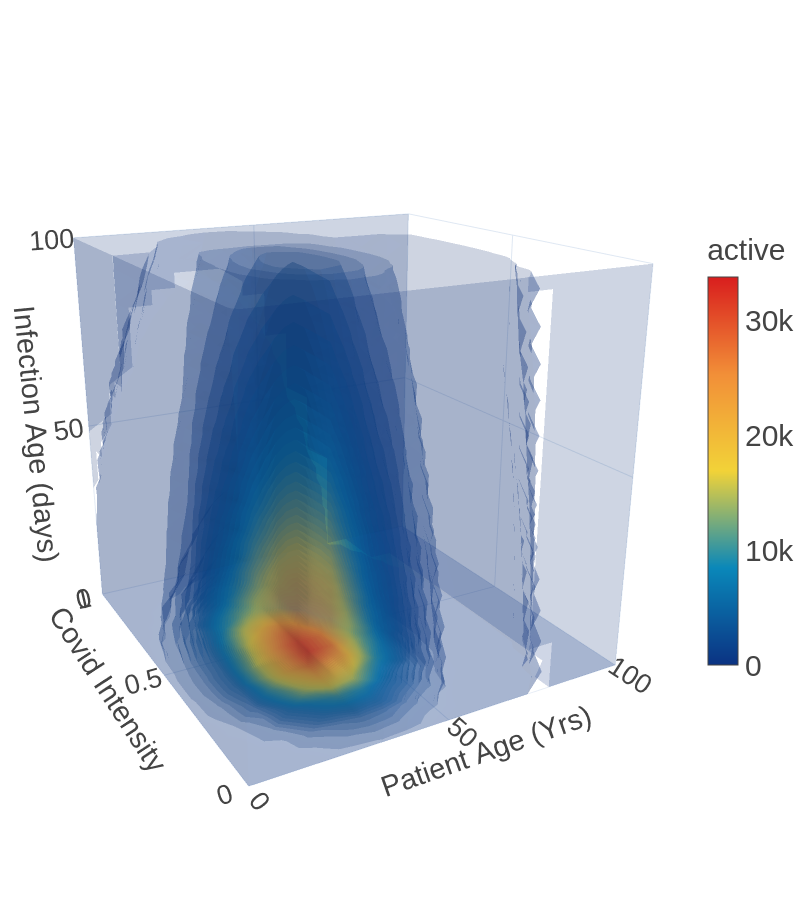}}}
\put(-6,13){\makebox(0,0){\includegraphics[clip, trim=0 0 0 200, width=5.5cm]{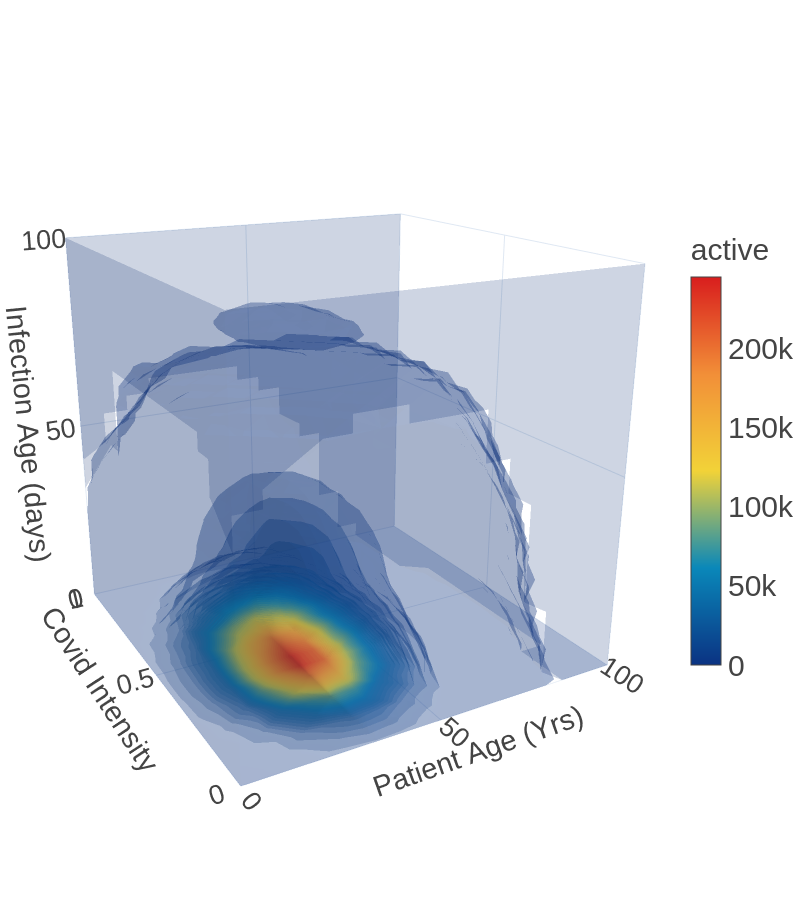}}}
\put(6,13){\makebox(0,0){\includegraphics[clip, trim=0 0 0 200,width=5.5cm]{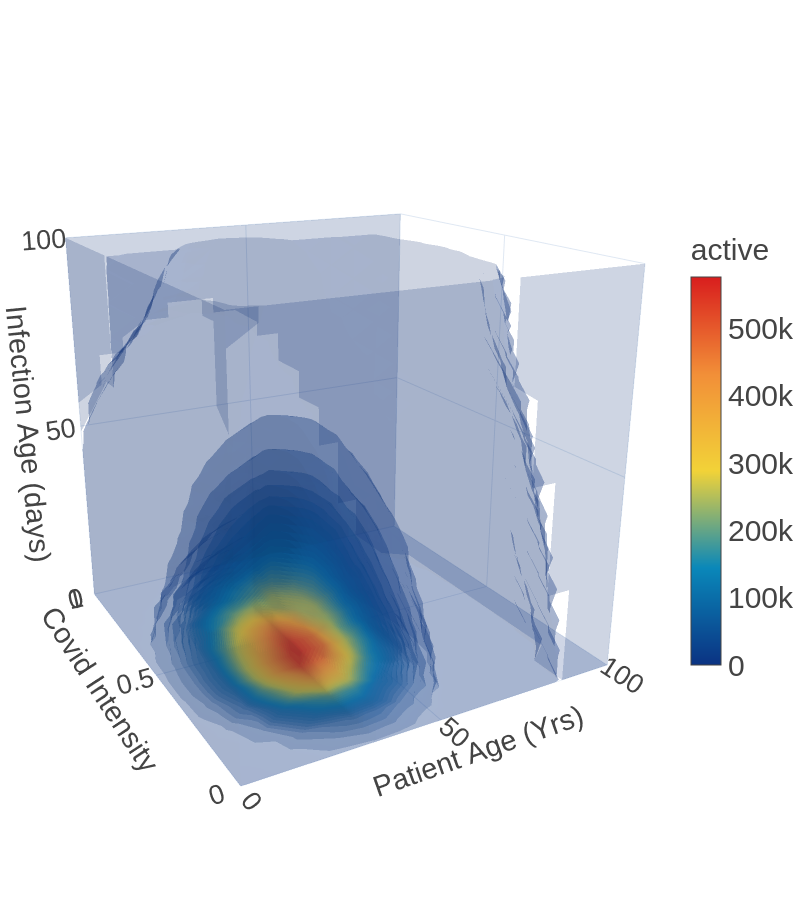}}}
\put(17.5,13){\makebox(0,0){\includegraphics[clip, trim=0 0 0 200,width=5.5cm]{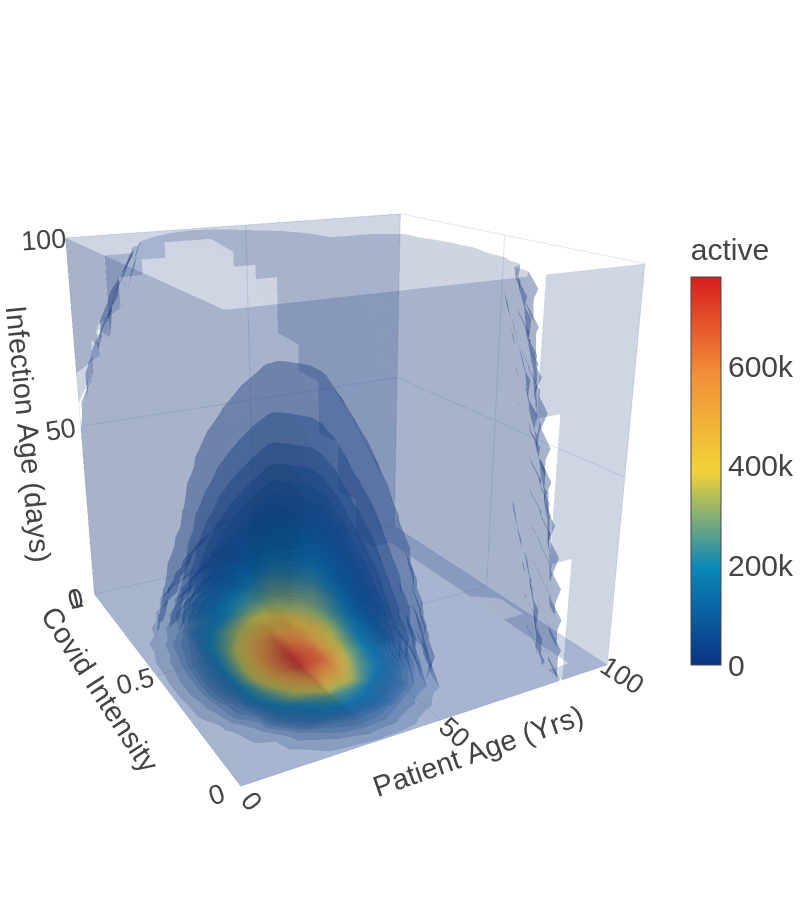}}}
\put(-8,18.25){(a)}
\put(3.5,18.25){(b)}
\put(15.5,18.25){(c)}
\put(-8, 8.){(d)}
\put(3.5, 8.){(e)}
\put(15.5, 8.){(f)}
\end{picture}
\end{center}
\caption{Distributions of Covid-19 population at different time instances, (a)~$t$=60, (b)~$t$=120, (c)~$t$=180, (d)~$t$=240, (e)~$t$=300 and (f)~$t$=365. \label{activeiso}}
\end{figure}
\begin{figure}[hb!]
\begin{center}
\unitlength5mm
\begin{picture}(12,18.5)
\put(-6,3){\makebox(0,0){\includegraphics[clip, trim=0 0 0 200,width=5.5cm]{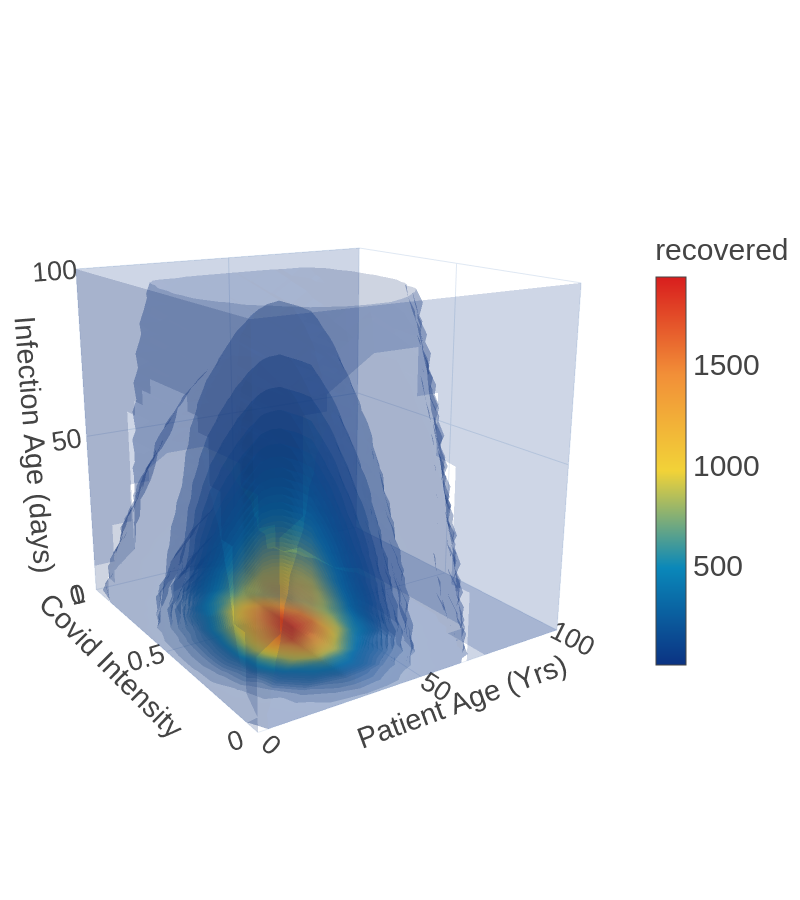}}}
\put(6,3){\makebox(0,0){\includegraphics[clip, trim=0 0 0 200,width=5.5cm]{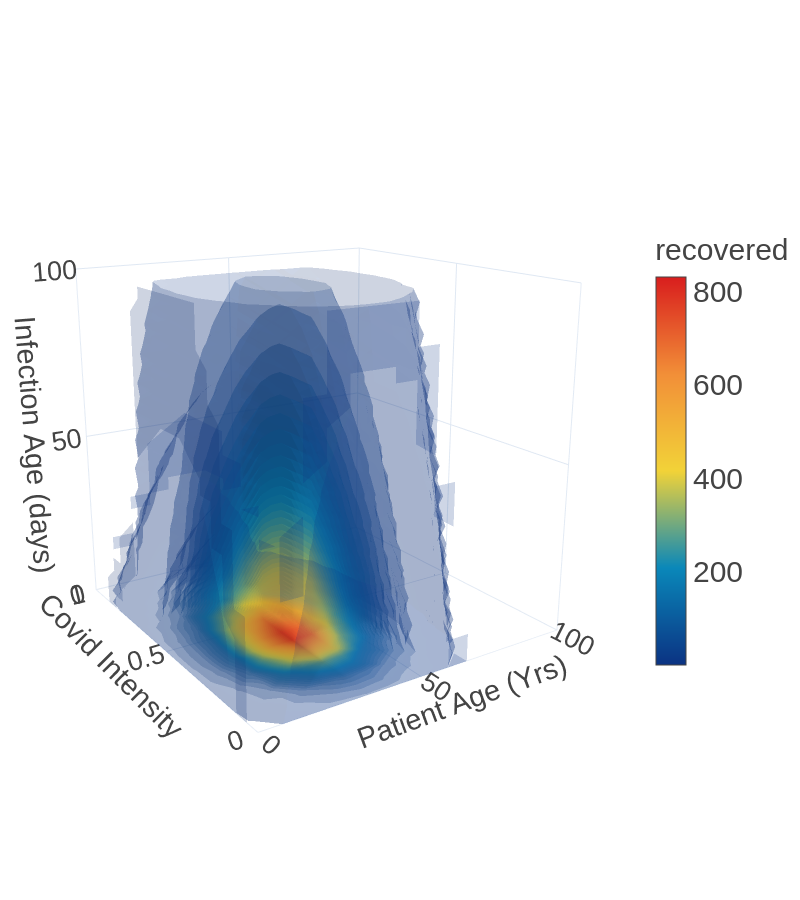}}}
\put(17.5,3){\makebox(0,0){\includegraphics[clip, trim=0 0 0 200,width=5.5cm]{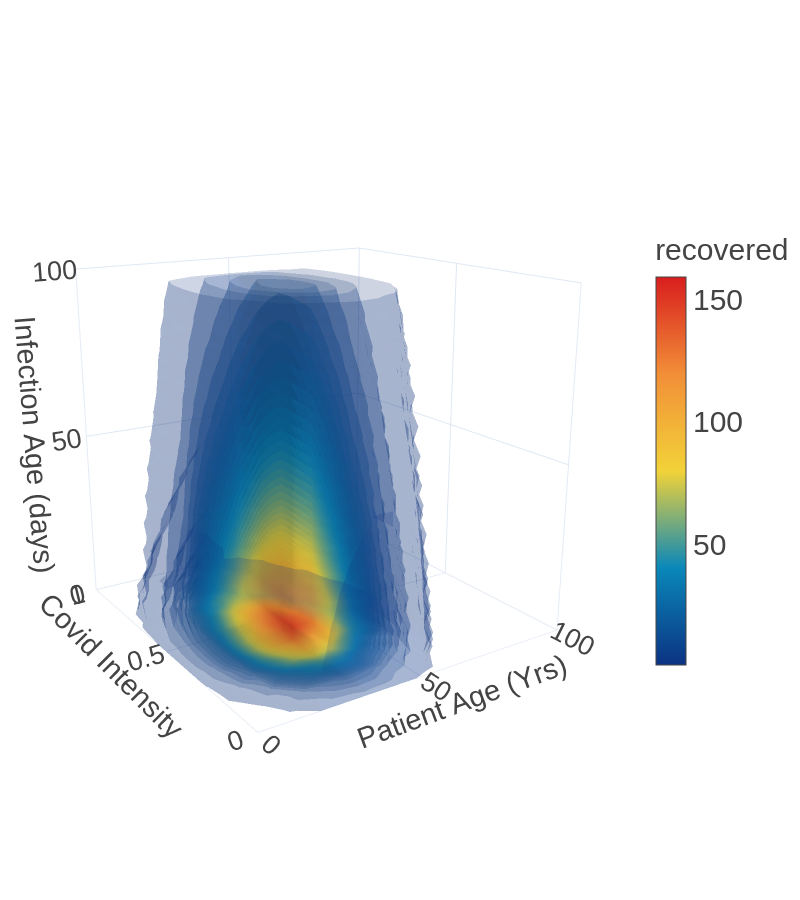}}}
\put(-6,13){\makebox(0,0){\includegraphics[clip, trim=0 0 0 200, width=5.5cm]{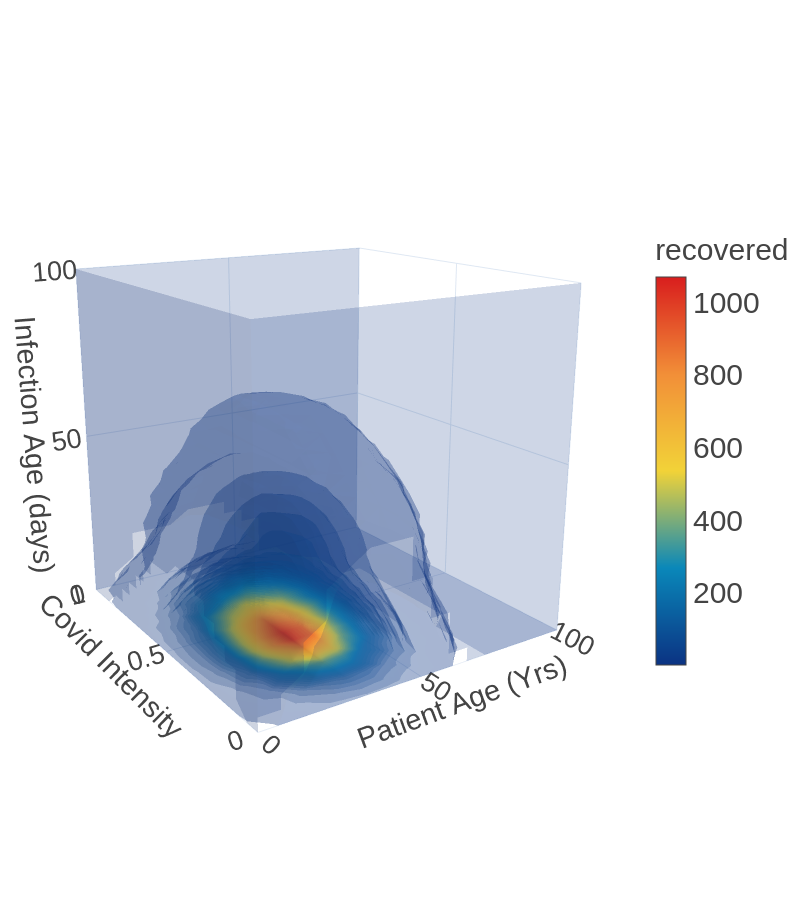}}}
\put(6,13){\makebox(0,0){\includegraphics[clip, trim=0 0 0 200,width=5.5cm]{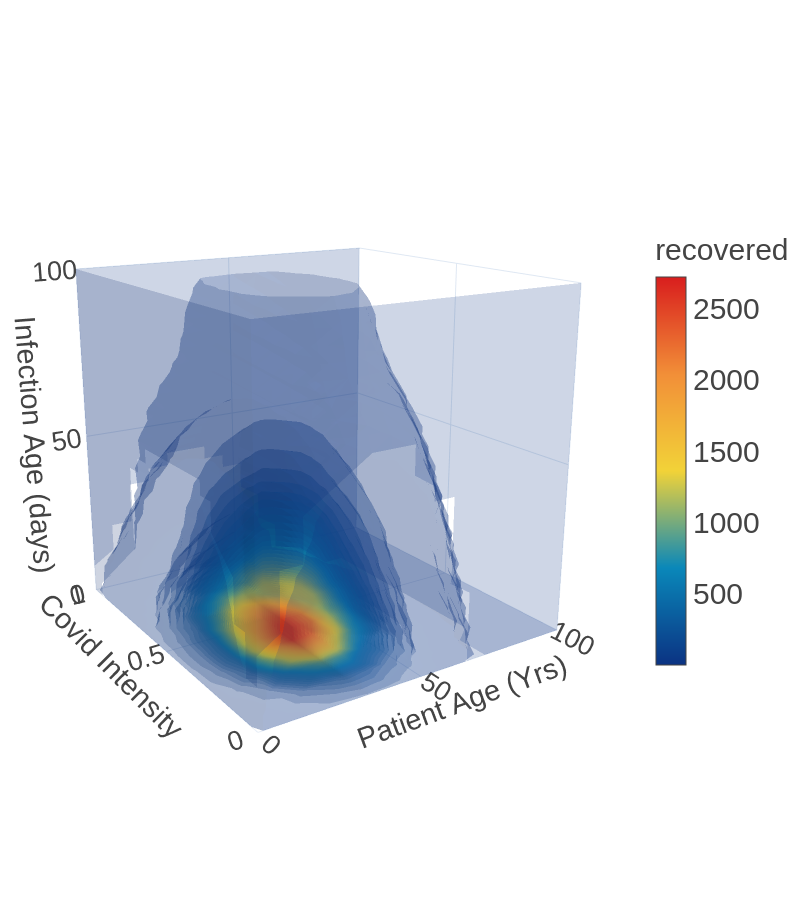}}}
\put(17.5,13){\makebox(0,0){\includegraphics[clip, trim=0 0 0 200,width=5.5cm]{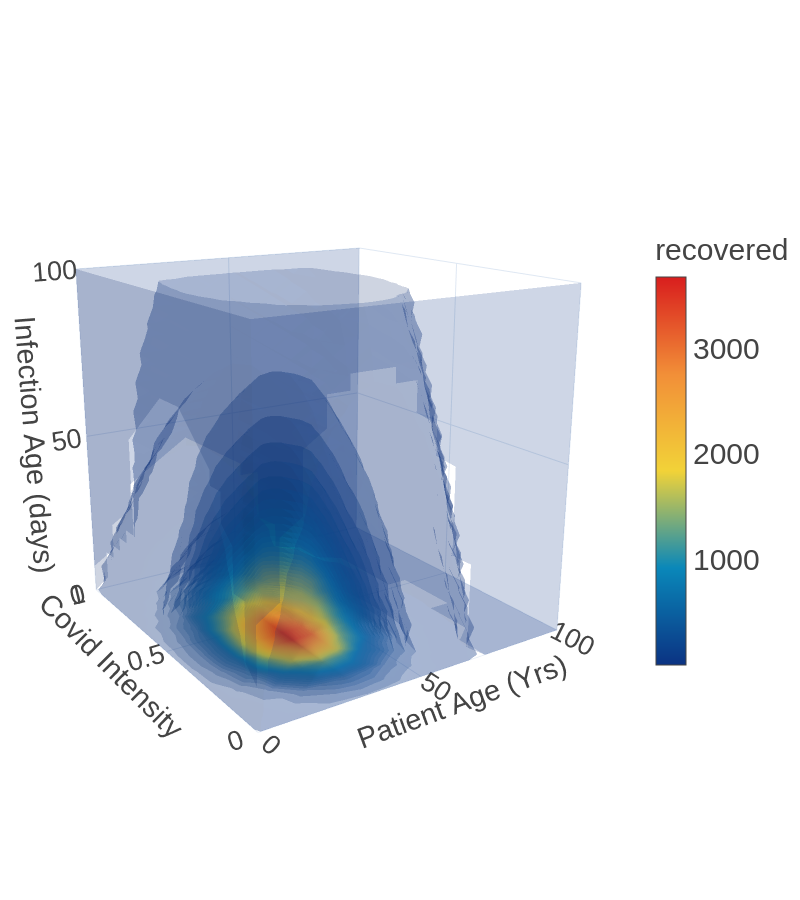}}}
\put(-8,18.25){(a)}
\put(3.5,18.25){(b)}
\put(15.5,18.25){(c)}
\put(-8, 8.){(d)}
\put(3.5, 8.){(e)}
\put(15.5, 8.){(f)}
\end{picture}
\end{center}
\caption{Distributions of recovered  population from Covid-19 at different time instances, (a)~$t$=60, (b)~$t$=120, (c)~$t$=180, (d)~$t$=240, (e)~$t$=300 and (f)~$t$=365. \label{recoviso}}
\end{figure}
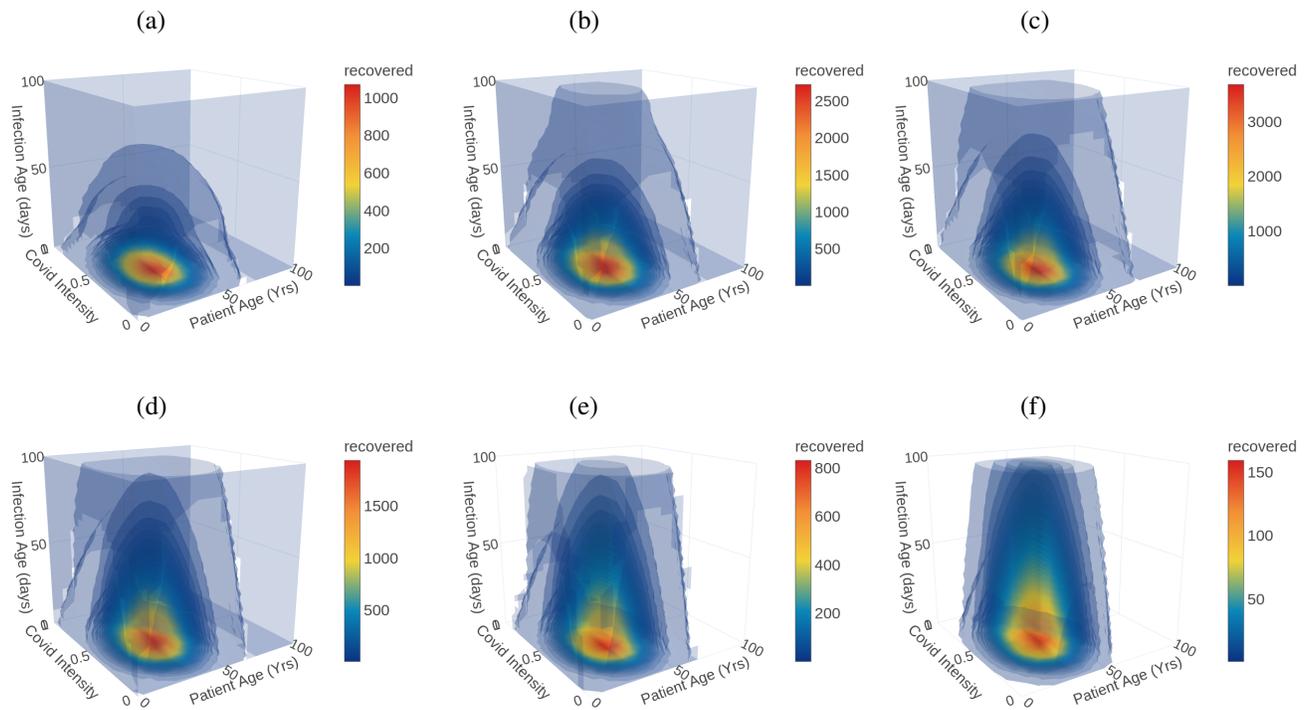

Our PBE model in fact predicts the distribution of the infected population over all the internal ordinates $\ell_v,~\ell_d,~\ell_a$. In the previous section, we have shown only the total number of infected, recovered and deceased populations. Now, to showcase one of the unique features of the model, we present and discuss the predicted population distribution for the Sunday lockdown scenario.  
Figure \ref{activeiso} shows the predicted distribution of active Covid-19 infected population over the ordinates $\ell_v$, $\ell_d$ and $\ell_a$ at different time instances (day 60, 120, 180, 240, 300 and 365) with their corresponding dates.

Predicting the severity of the infected population is crucial to plan the hospital requirements including antiviral treatments, quarantine, hospitalisation, ventilator support, and oxygen support. In particular, the information of asymptomatic and symptomatic infected population helps the policymakers to plan quarantine rules. Moreover, the death rate is a function of the severity of infection and is crucial to predict the causalities arising from the infection spread. 
The duration of infection plays a key role in epidemic modeling. Classical models usually assume a constant duration. However, the recovery and the death of the patient depend on the immunity, age and health of the patient, medical treatments etc and thus the duration of infection need not be a constant. In the proposed model, the duration of infection is considered as an independent internal ordinate. The recovered population distribution over the duration of infection and other internal ordinates for days 60, 120, 180, 240, 300 and 365 is shown in Figure \ref{recoviso} along with their corresponding dates. 
Crucially, the predicted distribution with duration of infection, especially at initial stages, is key to plan for testing and to make effective decisions on quarantine, hospitalization and discharging from hospitals. 
Finally, the age of the population is pivotal in epidemics like Covid-19 since it affects children and aged population severely. Therefore, it is incorporated into the proposed model as another independent ordinate. In fact, the newly infected population is added from the susceptible population across the age distribution through the nucleation term.
Moreover, the response to the antiviral treatment, death and recovery rates depend on age-specific health complications such as diabetics, cardiovascular disease, can also be incorporated in the PBE model with appropriate functions that depend on the age of the population.

\section*{Discussion}
Our spatio-temporal modeling framework is the first comprehensive partial differential equation model for predicting infectious disease spread. Computationally, our model is efficient compared to agent-based stochastic models. Mathematically, our PDE system is more compact and comprehensive compared to ODE-based compartmental models. Specifically, the PDE is a continuum description of the infected population whereas the compartmental models are a discrete representation. Crucially, in contrast to the existing models, our model provides an insight into the distribution of infected population (presented in previous sections). This information is important to plan policy interventions, especially in Covid-19 like pandemics. Not only prognostic estimates, but also diagnostic estimates for more detailed analysis using distribution can be performed with the proposed framework.

With more data and employing data-driven and machine learning approaches, we could further refine the parameters and functional forms of different model components to derive more insightful predictions. For example, to derive insights into the reopening of the workplace and educational institutions, the nucleation and advection vector could be modeled to account for interactions between different age groups and movement of people between homes and these places. The potential options for refining the model are virtually endless. In particular, there is no restriction on the choice of number of internal coordinates. For example, in addition to $\ell_v, \ell_d, \ell_a$, profession, mobility history, etc can also be added as internal coordinates. 

Even though we have emphasized Covid-19 pandemic in the present paper, the proposed model can readily be used for forecasting any other infectious disease spread. In future, a  data assimilative framework for a real-time update of forecasts can also be implemented.


\bibliography{lit}
 

\section*{Acknowledgements}
DS acknowledges the partial support from IISc Start-up grant, DST-INSPIRE Faculty Research Grant (04/2018/003591), and Arcot Ramachandran Young Investigator Award. SG acknowledges the support from SERB and DRDO for the grants that supported development of ParMooN. 

\section*{Author contributions statement}
Both authors contributed equally and reviewed the manuscript. 

\section*{Appendix}
As the focus of the present paper is to introduce the spatio-temporal predictive modeling framework for infectious disease spread, we use simple estimates and algebraic relations for certain parameters. Further, a few more assumptions are made on some parameters due to lack of actual data. Nevertheless, the features of the model and the insights into the prediction of the spread can be seen even with these assumptions. 

\subsection*{Numerical Scheme}
 Let $T_\infty=365$ days, $d_\infty=400$ days, $a_\infty=45,625$ days  and $\Omega_{\rm x}:=\cup\Omega_k$, $k=1,\ldots,M$, where $M$ being the   number of states and union territories in India. We assume that there will be no inter- or intra-state and international movements, that is,  $\bu = 0 $,  no growth in level of infection, that is, $G_{\ell_v}=0$, the growth age is negligible and no source. Nevertheless, population distribution in all three internal ordinates $\ell_v$, $\ell_d$ and $\ell_a$ are described and tracked. Hence, the PBE in model~\eqref{model} becomes
\begin{equation}\label{lockmodel}
 \ds\frac{\partial I}{\partial t}     +   \frac{\partial I}{\partial \ell_d}  +C I  = 0  \quad  \rm{ in }  \quad  (0,T_\infty]\times\Omega_{x}\times\Omega_\ell.
\end{equation}

\subsubsection*{Operator splitting finite element scheme}
A finite element scheme~\cite{GTBook16} based on operator splitting~\cite{GAN13,GT11a,GANW10} is used to solve the high-dimensional PBE model~\eqref{lockmodel}.  Applying, operator splitting to~\eqref{lockmodel}, we get 
\\ 

\noindent \textbf{Step 1. ($x$-direction)}
For given $I(t^a,x,\ell)$ with $I(t^a=0,x,\ell)=I_0$, find $\tilde I(t^b,x,\ell)$ in $(t^a , t^b)$ for all $\ell\in\Omega_\ell$ such that
\begin{equation}\label{x-lockmodel}
 \begin{array}{l}
 \ds\frac{\partial \tilde I}{\partial t} + C I   = 0,\quad
  \tilde I(t^a,x,\ell) = I(t^a,x,\ell)   \quad \rm{ in }  \quad  \Omega_{x}  \,,
\end{array}
\end{equation}
\noindent \textbf{Step 2. ($\ell_d$-direction)}
For given $\tilde I(t^b,x,\ell)$, find $  I(t^b,x,\ell)$ in $(t^a , t^b)$  for all $x\in\Omega_{\rm x}$   $\ell_v \in L_v$ and  $\ell_a \in L_a$ such that
\begin{equation}\label{s2-lockmodel}
 \begin{array}{l}
 \ds\frac{\partial  I}{\partial t}   +  \frac{\partial  I}{\partial \ell_d}    = 0,\quad I(t^a,x,\ell) = \tilde I(t^b,x,\ell) \quad \rm{ in }  \quad   \Omega_\ell \,;\qquad
   I(t,\ell_v, 0, \ell_a)  =B_{nuc} \quad   \rm{ in }  \quad   L_a ,\\
  
\end{array}
\end{equation}
In the $x$-direction, the evaluation equation~\eqref{x-lockmodel} has to be solved for every $\ell\in\Omega_\ell$ by considering $\ell$ as a parameter. Similarly, the system \eqref{s2-lockmodel}  has to be solved in $\ell_d$-direction for every $x\in\Omega_{\rm x}$   $\ell_v \in L_v$ and  $\ell_a \in L_a$ by considering these variables as parameters. The backward Euler   and discontinues Galerkin with upwind  methods are used for the temporal and the spatial discretisation, respectively. The implementation of the splitting algorithm in the finite element context has been presented in these papers~\cite{GAN13,GT11a,GANW10}.

\subsubsection*{Parameters for  Covid-19  predictions}
The nucleation  model $B_{\textrm nuc}$  defined in \eqref{eq:Bnuc_eq2} is considered with $R_0=3.35$, $f_1=1$, $f_2=1$, $f_4=1$ and
 \[
f_3(t,S_D) = 1. -  1./(1.+ \exp(-(S_D(t) - 0.5)/0.1)), \quad
     S_D(t) = 
        \left\{ 
            \begin{array}{lc}
               0.7 +  0.001333~t & 0\leq t<15  \,\\
               0.72 + 0.004285(t-15) & 15\leq t<36  \,\\  
                0.81 - 0.004(t-36) & 36\leq t<51  \,\\     
                0.75 + 0.0012(t-51.) & 51\leq t<72  \,\\                   
                0.3 + ds(t-72) & \textrm{else}
            \end{array}  
        \right.  
\]
Moreover, the values given in Table~\ref{dsvalues} are used for $ds$ and $f_1$ to perform scenario analysis. 
 \begingroup
\setlength{\tabcolsep}{10pt} 
\renewcommand{\arraystretch}{1.25} 
\begin{table}[h!] 
\centering
\begin{tabular}{|c|c|c|l|}
 \hline  
  Scenarios  & Current Trend & Better Trend & Periodic lockdowns\\ \hline  
  $ds$  &  0.000333  & 0.0005 & 0.000333 \& \\
  & &  & $f_1 = 0.01$ on lockdown days \\ \hline  
\end{tabular}
\caption{Parameter values used in scenario analysis.}
\label{dsvalues}
\end{table}
\endgroup

Furthermore, the quarantine function 
\[
\gamma_Q(\ell_d) = 
\left\{ 
    \begin{array}{lc}
       1 & 0\leq t<1  \,\\
       0.9(1./(1.+ \exp(-(l_v-0.4)/0.1)))*(1./(1.+ \exp(-(l_d-5.1)/2)))   & \textrm{else,}
    \end{array}  
\right.  
\]
is used in all scenarios.
Finally, the recovery and death rate functions are fitted as 
\[
C_{ID}(t,\ell_v) = 
\left\{ 
    \begin{array}{ll}
       0 & \ell_v<0.64  \,\\
        0.0475 - 0.000357~t   & \ell_v\geq 0.64 \textrm{ and }  t<21 \\
       0.0475 + 0.000208(t-21)  & \ell_v\geq 0.64 \textrm{ and }  21\leq t<36 \\        
       0.0475  & \textrm{else,}
    \end{array}  
\right., \quad
C_{R}(t) = 
\left\{ 
    \begin{array}{ll}
        0.01 + t*0.00058 & t<65  \,\\
        0.0475  & \textrm{else.}
    \end{array}  
\right.
\]

\noindent\textbf{Remark:} A factor $N_S(t)/N(t)$ needs to be introduced in $B_{nuc}$  when herd immunity  develops.

\subsubsection*{Initial infection number density}
 For India, the data to estimate $N_D(\mathbf{x})$ is downloaded from a publicly sourced database \cite{covid19tracker}.
 To distribute the number density among the internal ordinates, we employ distribution fits as given in equation~\eqref{eq:I0dist}. Data set of age downloaded from Statista \cite{statista} is used to fit $a_1,b_1,c_1$ of equation \eqref{eq:I0dist}. 
  


\end{document}